\documentclass[showpacs,superscriptaddress,preprintnumbers,amsmath,amssymb,fleqn]{revtex4-1} 
\usepackage{graphicx}
\usepackage{dcolumn}
\usepackage{bm}

\begin{document}

\preprint{INHA-NTG-01/2011}

\title{A phenomenological description on an incoherent Fermi
liquid near optimal doping in high T$_{c}$ cuprates}

\author{Ki-Seok Kim}
\email[E-mail: ]{kimks@kias.re.kr}
\affiliation{Asia Pacific Center for Theoretical Physics,
Pohang, Gyeongbuk 790-784, Republic of Korea}
\affiliation{Department of Physics, Pohang University of Science and
Technology, Pohang, Gyeongbuk 790-784, Korea}

\author{Hyun-Chul Kim}
\email[E-mail: ]{hchkim@inha.ac.kr}
\affiliation{Department of Physics, Inha University, Incheon 402-751
 Korea} 
\affiliation{School of Physics, Korea Institute for Advanced Study,
  Seoul 130-722,  Korea} 
\date{\today}

\begin{abstract}
Marginal Fermi-liquid physics near optimal doping in high T$_{c}$
cuprates has been explained within two competing scenarios such as
the spin-fluctuation theory based on an itinerant picture and the
slave-particle approach based on a localized picture. In this
study we propose an alternative scenario for the anomalous
transport within the context of the slave-particle approach.
Although the marginal Fermi-liquid phenomenology was interpreted
previously within deconfinement of the compact gauge theory,
referred to as the strange metal phase, we start from confinement, 
introducing the Polyakov-loop parameter into an SU(2) gauge theory
formulation of the t-J model. The Polyakov-loop parameter gives
rise to incoherent electrons through the confinement of spinons
and holons, which result from huge imaginary parts of self-energy
corrections for spinons and holons. This confinement scenario
serves a novel mechanism for the marginal Fermi-liquid transport
in the respect that the scattering source has nothing to do with
symmetry breaking. Furthermore, the incoherent Fermi-liquid state
evolves into the Fermi liquid phase through crossover instead of
an artificial second-order transition as temperature is lowered,
where the crossover phenomenon does not result from the
Anderson-Higgs mechanism but originate from an energy scale in the
holon sector. We fit an experimental data for the electrical
resistivity around the optimal doping and find a reasonable match
between our theory and the experiment.
\end{abstract}


\maketitle

\section{Introduction}
The scattering mechanism for anomalous transport near optimal
doping in high T$_{c}$ cuprates has been a long standing puzzle
since the discovery of high T$_{c}$ superconductivity
\cite{Optimal_Doping_Review}. A phenomenological description was
proposed immediately, referred to as the marginal Fermi-liquid
Ansatz, where the existence of critical fluctuations is essential
\cite{Varma_MFL_Review}. In particular, the nature of such
critical excitations is suggested to be local, where dominant
contributions for scattering with electrons result from their
frequency dependence. As a result, dynamics of electrons becomes
incoherent, described by the local self-energy correction linearly
proportional to frequency/temperature for its imaginary part. The
relaxation rate can be identified with the scattering rate
associated with transport because vertex corrections for
relaxation of currents vanish automatically due to the locality of
critical fluctuations. This explains the origin of the
quasi-linear behavior of the electrical resistivity in temperature
near the optimally doped high T$_{c}$ cuprates. Later, a
microscopic description for the marginal Fermi-liquid
phenomenology was developed, called the circulating
orbital-current model based on the three-band Hubbard model
\cite{Varma_current_model}.

A standard description for such an incoherent Fermi-liquid state
is based on the spin-fluctuation model, strongly
correlated Fermi liquid being illustrated in the vicinity of an
itinerant antiferromagnetic quantum critical
point~\cite{Chubukov_AFQCP_Review}. Unfortunately, the low-energy 
physics of this effective field theory is not known even in the
large-$N$ limit, where $N$ is the spin degeneracy and $1/N$ is the
only expansion parameter \cite{SungSik_Genus}. In spite of
uncontrolled approximation, the quasi-$T$-linear electrical
resistivity was demonstrated, the results of which was in remarkable 
agreement with experimental data near optimal
doping~\cite{Chubukov_AFQCP_Review}. The main difficulty of this
Fermi-liquid based approach is that the connection to an
insulating-like state in the underdoped region is not clear.

Another popular approach is based on the slave-particle
representation, where an electron field is assumed to be a
composite operator given by a fermionic spinon field and a bosonic 
holon field, carrying spin and charge quantum numbers
separatively~\cite{Lee_Nagaosa_Wen}. The spin-charge separation
scenario being followed, the anomalous transport property was
explained in terms of the incoherent dynamics of holons, where the
scattering mechanism for the incoherent dynamics is served from spin  
chirality fluctuations, described by gauge fluctuations in this
effective field theory~\cite{SM_U1GT}. The spin-charge separated 
anomalous metallic state evolves into the Fermi liquid phase via
the condensation of holons which results in coherent electrons. An 
essential aspect is that this scenario is based on a paramagnetic
Mott insulating phase called a spin liquid
state~\cite{Spin_Liquid}. As a result, the evolution of electron 
dynamics from a Mott insulator to a Fermi liquid can be
systematically described. However, the Anderson-Higgs mechanism
gives rise to an artificial transition from the finite-temperature
spin liquid to the Fermi liquid instead of the crossover, which
remains as an open issue. Furthermore, the connection between the
slave-particle approach and the spin-fluctuation scenario is also
elusive near optimal doping.

In this study, we propose an alternative scenario to describe the 
marginal Fermi-liquid phenomenology near optimal doping and to
resolve the artificial finite-temperature transition to the Fermi
liquid. Although the spin-charge separation is assumed in the
underdoped region of high $T_{c}$ cuprates~\cite{Lee_Nagaosa_Wen}, 
spinons and holons are conjectured to be confined around the
optimal doping region. Based on this confinement Ansatz, we
reproduce the experimental data for the electrical
transport~\cite{Data}, which implies that the confinment of the
spinons and holons may play an essential role in dynamics of
incoherent electrons. In addition to the non-Fermi liquid physics
via the Polyakov-loop parameter, we realize the crossover instead
of the second order transition, described by the coherence of such
confined electrons.

The confinement of the spinons and holons will be described
phenomenologically, the Polyakov-loop parameter being
introduced~\cite{Ployakov_Loop} to measure an effective fugacity
of such fractionalized excitations. The Polyakov-loop parameter
becomes condensed to allow the deconfinement at high temperatures
while it vanishes to cause the confinement at low temperatures. The
crossover from the marginal Fermi liquid to the Fermi liquid is
conjectured to occur below the confinement-deconfinement crossover
temperature. In this paper we do not discuss how the spin liquid state
in the underdoped region evolves into the Fermi liquid phase in the
overdoped region, leaving it for more serious future
investigations beyond this study.

The Nambu-Jona-Lasinio model (NJL) has been successfully used as
an effective model for nonperturbative Quantum Chromodynamics
(QCD) in explaining low-energy phenomena such as properties of
hadrons, since it describes successfully the spontaneous breakdown
of chiral symmetry
(SB$\chi$S)~\cite{NJL_review1,NJL_review2,NJL_review3}. The model 
consists of only quark degrees of freedom, the gluons being
being integrated out, for example, via
instantons~\cite{SchaeferShuryak,Diakonov:2002fq}. While the NJL
model explains certain aspects of the QCD vacuum and spectra of
lowest-lying hadrons, it does not contain any feature of the
confinement. One of the simplest ways to bring this confinement
back into this framework is to introduce an additional
temporal static gluon field, which can be coupled to quarks. This
effective model is called the Polyakov extended NJL model
(PNJL)~\cite{Fukushima}. The thermally averaged Polyakov loop can
be taken as an order parameter for deconfinement in the limit of
the heavy-quark mass. In this case, the crossover of
confinement-deconfinement is featured by the spontaneous breakdown
of the $Z_{N_c}$ center symmetry with the number of color
$N_c$~\cite{Polyakov,Susskind,McLerran,Svetitsky}. The presence of
the dynamical quark breaks explicitly this symmetry, so that the
Polyakov loop is no more order parameter. However, it was shown
that the Polyakov loop describes a rapid crossover in the vicinity
of the critical point of the deconfinement phase, which indicates
that we still can consider the Polyakov loop as a signature for
the deconfinement. The results of various susceptibilities from
the PNJL are known to be consistent with the lattice
simulation~\cite{PNJL_Models}. In addition, the PNJL scheme has
been employed for the case of finite density of
quarks~\cite{PNJL_Density} to study a generic phase diagram in
quark matter. Analogously, the PNJL will provide a good
phenomenological framework for investigating the crossover from
non-Fermi liquid to Fermi liquid.

We point out the previous study in which the spinon-holon binding was
introduced in the U(1) slave-boson approach~\cite{TKNG}. Although it
is in line with the present work, the spinon-holon bound state is
distinguished from the spinon-holon confined state: such a
fractionalized particle is a well-defined excitation above the
binding energy in the previous study~\cite{TKNG}, while it is
unstable for all energy scales due to the presence of the
Polyakov-loop parameter in our formulation.

The present paper is organized as follows: In Section II, we
briefly review the PNJL model in the context of an SU(2) gauge
theory of the t-J model. In Section III, we describe the coherent
crossover from non-Fermi liquid to Fermi liquid with the effects
of the confinement considered via the Polyakov-loop parameter. We
present the results of the electrical resistivity which shows the
crossover at low temperatures. The final Section is devoted the
summary and discussion of the present results.

\section{A PNJL-model description for an SU(2) gauge theory
of the t-J model}

\subsection{Review on an SU(2) slave-boson theory of the t-J model} 

We start from the t-J model for doped Mott insulators
\begin{equation}
  \label{eq:1}
H = - t \sum_{\langle i j \rangle} (c_{i\sigma}^{\dagger}c_{j\sigma} +
H.c.) + J \sum_{\langle i j \rangle} (\vec{S}_{i}\cdot\vec{S}_{j}
- \frac{1}{4}n_{i}n_{j})\, ,
\end{equation}
where hopping of electrons is allowed only when the site for hopping
is empty. This constraint raises an important problem for the
quantization because the electron operator with it does not satisfy
the anti-commutation relation. One way to avoid this problem is to
write an electron field as a composite operator
\begin{eqnarray}
  \label{eq:comp}
c_{i\uparrow} &=& \frac{1}{\sqrt{2}} h_{i}^{\dagger}
\psi_{i\uparrow} = \frac{1}{\sqrt{2}} (b_{i1}^{\dagger}f_{i1} +
b_{i2}^{\dagger}f_{i2}^{\dagger}) , \cr
c_{i\downarrow} &=&
\frac{1}{\sqrt{2}} h_{i}^{\dagger} \psi_{i\downarrow} =
\frac{1}{\sqrt{2}} (b_{i1}^{\dagger}f_{i2} -
b_{i2}^{\dagger}f_{i1}^{\dagger})\, ,
\end{eqnarray}
where the holon and spinon fields are expressed, respectively, as
\begin{equation}
h_{i} \;=\;  \left( \begin{array}{c} b_{i1} \\ b_{i2}
\end{array}\right),\;\;\;\;\;
\psi_{i\sigma} = \left(
\begin{array}{c} f_{i\sigma} \\ \epsilon_{\sigma\sigma'}
  f_{i\sigma'}^{\dagger} \end{array}\right).
\label{eq:holon_spinon}
\end{equation}
$\epsilon_{\sigma\sigma'}$ is an antisymmetric tensor, i.e.,
$\epsilon_{12} = + 1$ and $\epsilon_{21} = -
1$. Equation~(\ref{eq:comp}) is known to be the SU(2) slave-boson
representation, originally introduced in Ref.~\cite{SU2SB}, which
incorporates low-lying fluctuations missed in the U(1) slave-boson  
representation of the t-J model near
half-filling~\cite{Comment_SU2SB}. The holon and spinon fields carry   
respectively charge and spin quantum numbers, satisfying the
commutation and anti-commutation relations, respectively. Since
this representation enlarges the physical Hilbert space, a
constraint for the single occupancy should be introduced as
follows:
\begin{equation}
\frac{1}{2}
\psi_{i\alpha}^{\dagger} \tau_{k} \psi_{i\alpha} + h_{i}^{\dagger}
\tau_{k} h_{i} = 0\, ,
\label{eq:constraint}
\end{equation}
which recovers the physical Hilbert space and satisfies the
electron anti-commutation relation. The $\tau_{k}$ is the Pauli matrix  
with $k = 1, 2, 3$, which acts on the SU(2) spinor space of
Eq.~(\ref{eq:holon_spinon}).

Using this SU(2) slave-boson representation, one can rewrite the t-J
model for electrons as an effective one for holons and
spinons~\cite{Lee_Nagaosa_Wen} 
\begin{equation}
Z \;=\; \int D \psi_{i\alpha} D h_{i} D U_{ij} D a_{i0}^{k} e^{-
\int_{0}^{\beta} d \tau L}
\end{equation}
with the Lagrangian
\begin{eqnarray}
L &=& \frac{1}{2} \sum_{i}
\psi_{i\alpha}^{\dagger} (\partial_{\tau} - i
a_{i0}^{k}\tau_{k})\psi_{i\alpha} + J \sum_{\langle i j \rangle} (
\psi_{i\alpha}^{\dagger}U_{ij}\psi_{j\alpha} + H.c.) +
\sum_{i} h_{i}^{\dagger}(\partial_{\tau} - \mu  - i
a_{i0}^{k}\tau_{k}) h_{i} \cr && +\; t \sum_{\langle i j \rangle} (
h_{i}^{\dagger}U_{ij}h_{j} + H.c.) + J \sum_{\langle i j \rangle}
\mathrm{tr}[U_{ij}^{\dagger}U_{ij}],
\label{eq:Lag}
\end{eqnarray}
where $a_{i0}^{k}$ is a Lagrange multiplier field to impose the
SU(2) slave-boson constraint given in Eq.~(\ref{eq:constraint}),
identified with the time component of the SU(2) gauge field. The
matrix of the order parameter can be written as
\begin{equation}
  \label{eq:3}
U_{ij} = \left( \begin{array}{cc} - \chi_{ij}^{\dagger} & \Delta_{ij}
    \\ 
\Delta_{ij}^{\dagger} & \chi_{ij} \end{array}\right)\,,
\end{equation}
which comes from the standard bosonization procedure for
interactions, where the spin-singlet interaction channel of the
J-term is decomposed into particle-hole exchange hopping and
particle-particle d-wave pairing scattering channels. The
Hubbard-Stratonovich transformation is performed to introduce
effective bosonic order parameters, $\chi_{ij}$ and $\Delta_{ij}$
associated with the particle-hole and particle-particle channels,
respectively. Such order parameter fields are given by $\chi_{ij}
= \langle f_{i\sigma}^{\dagger}f_{j\sigma} + H.c. \rangle$ and
$\Delta_{ij} = \langle \epsilon_{\sigma\sigma'} f_{i\sigma}
f_{j\sigma'} \rangle$, respectively, in the saddle-point
approximation. We want to emphasize that Eq.~(\ref{eq:Lag}) can be
regarded as one reformulation of the t-J model, decomposing an
electron field into the spinon and holon fields as in
Eq.~(\ref{eq:comp}), where the Gutzwiller projection, i.e., the
constraint for hopping, is replaced with the exact integration
over $a_{i0}^{k}$.

Employing the mean-field approximation for $U_{ij}$ and
$a_{i0}^{k}$, Wen and Lee~\cite{Lee_Nagaosa_Wen} found the phase
diagram of the effective theory represented by Eq.~(\ref{eq:Lag})
in the $(\delta, T)$ plane with a fixed $J/t$, where $\delta$
denotes the hole concentration and $T$ stands for the temperature.
The optimally hole-doped region at high temperatures is described
by $U_{ij}^{SM} = - i \chi I$ and $\langle h_{i} \rangle = 0$ with
$a_{i0}^{k} = 0$, which is called the strange metal (SM) phase,
where the spinons form a large Fermi surface but the incoherent
electron spectra are only observed~\cite{Underdoped_Phase}. The
Fermi liquid state appears from the SM phase, given by
the condensation of holons $\langle b_{i1} \rangle \not= 0$ and 
$\langle b_{i2} \rangle = 0$, which results from the change of the
chemical potential $i a_{i0}^{k} \not= 0$.

Low-energy physics and the stability of each phase should be
investigated beyond the mean-field description, quantum
fluctuations being introduced and an effective field theory being
constructed. Considering quantum fluctuations $U_{ij}^{SM} = - i
\chi e^{i a_{ij}^{k} \tau_{k} }$ in the SM phase, we can discuss
its low-energy physics based on the following SU(2) gauge theory
\cite{SF_vs_SM}
\begin{eqnarray}
{\cal L}_{\mathrm{eff}} &=& \psi_{\alpha}^{\dagger} (\partial_{\tau} -
\mu_{s}\tau_{3} - ia_{\tau}^{k} \tau_{k} ) \psi_{\alpha} +
\frac{1}{2m_{\psi}} |(\partial_{i} - i a_{i}^{k} \tau_{k}
)\psi_{\alpha}|^{2} \cr &+& h^{\dagger}(\partial_{\tau} - \mu -
ia_{\tau}^{k}\tau_{k})h + \frac{1}{2m_{h}} |(\partial_{i} -
ia_{i}^{k}\tau_{k})h|^{2} +
\frac{1}{4g^{2}}[\partial_{\mu} a_{\nu}^{k} -
\partial_{\nu} a_{\mu}^{k} - g \epsilon_{klm} a_{\mu}^{l}
a_{\nu}^{m}]^{2}\,,
\end{eqnarray}
where the temporal $a_{\tau}^{k}$ and spatial $a_{i}^{k}$
components of the SU(2) gauge fields come from the Lagrange
multiplier field $a_{i0}^{k}$ and phase of the order parameter
matrix $a_{ij}^{k}$, respectively. The $\mu_{s}$ represents a
spinon chemical potential, which originates from the vacuum
expectation value of the time-component gauge field, as discussed
just above. While $m_{\psi} \propto 1/J$ is a band mass for the
spinons, $m_{h} \propto 1/t$ is that for the holons. The $g$
denotes an effective coupling constant between the spinons
(holons) and the gauge bosons. Note that the SU(2) Maxwell
dynamics results from high energy fluctuations of the spinons and
holons. In this effective field theory the spinons interact with
the holons via SU(2) gauge fluctuations. Such gauge interactions
have been proposed as the source for the anomalous transport in
the SM phase~\cite{SM_U1GT}. However, it is not enough to treat
gauge fluctuations perturbatively in order to simulate the
Gutzwiller projection~\cite{Gutzwiller_tJ,Mudry_Confinement}.
Moreover, such an approach based on the deconfinement cannot
recover electron excitations at low temperatures without the
Anderson-Higgs mechanism, which means that the Fermi liquid phase
is never obtained without the artificial transition. Thus, we
propose in the present work one possible way of describing the
crossover from the SM to the Fermi liquid around the optimal
doping, based on the confinement-deconfinement feature.

\subsection{PNJL model in the strange metal phase}
We write down the covariant derivative as
\begin{equation}
D_{\mu} = \partial_{\mu} - i \phi \tau_{3} \delta_{\mu \tau} -
ia_{\mu}^{k} \tau_{k} ,
\end{equation}
where $\phi$ corresponds to a nontrivial vacuum contribution,
nothing but the mean-field part of the gauge field associated with
the Polyakov-loop parameter, and $a_{\mu}^{k}$ express quantum
fluctuations around the nontrivial vacuum state. Integrating over
such quantum fluctuations, we find an effective Lagrangian of the PNJL 
decomposed into the matter and gauge parts
\begin{equation}
\mathcal{L}_{\mathrm{PNJL}} = \mathcal{L}_{\mathrm{PNJL}}^{M} +
\mathcal{L}_{\mathrm{PNJL}}^{G}\, ,
\end{equation}
where $\mathcal{L}_{\mathrm{PNJL}}^{M}$ and
$\mathcal{L}_{\mathrm{PNJL}}^{G}$ describe matter and gauge sectors,
respectively. The matter part is written as
\begin{equation}
\mathcal{ L}_{\mathrm{PNJL}}^{M} =
\psi_{\alpha}^{\dagger} (\partial_{\tau} - i \phi \tau_{3} -
\mu_{s} \tau_{3}) \psi_{\alpha} + \frac{1}{2m_{\psi}} |
\partial_{i} \psi_{\alpha}|^{2} +
h^{\dagger}(\partial_{\tau} - i \phi \tau_{3} - \mu )h +
\frac{1}{2m_{h}} | \partial_{i} h|^{2} + \mathcal{L}_{\mathrm{int}}\, ,
\label{eq:matter}
\end{equation}
where $\mathcal{L}_{\mathrm{int}}$ is generated from the gauge
interactions $a_{\mu}^{k}$, which is assumed to be local as follows:
\begin{equation}
  \label{eq:8}
\mathcal{L}_{\mathrm{int}} = g_{I} \Bigl(\psi_{\alpha}^{\dagger} \tau_{k}
\psi_{\alpha} + h^{\dagger} \tau_{k} h \Bigr)^{2} + g_{J}
\Bigl(\frac{1}{2m_{\psi}} [\psi_{\alpha}^{\dagger} \tau_{k}
\partial_{i} \psi_{\alpha} - H.c.] + \frac{1}{2m_{h}} [h^{\dagger}
\tau_{k} \partial_{i} h - H.c.]\Bigr)^{2} .
\end{equation}
The $g_{I}$ and $g_J$ denote an effective coupling strength for
isospin density interactions and isospin current ones,
respectively. This local approximation is well utilized in the QCD
context, realizing SB$\chi$S
successfully~\cite{PNJL_Models}. Introducing non-local
interactions will improve the qualitative picture more
quantitatively. However, we will assume local interactions, since our
goal is to show the emergence of the confinement-deconfinement
crossover instead of the precise description of chiral symmetry
breaking. The local current-current interactions are expected to be
irrelevant in the renormalization group sense, so that they will be
neglected for simplicity. Using the following identity $\tau_{nm}^{k}
\tau_{pq}^{k} = - \delta_{nm} \delta_{pq} + 2 \delta_{nq}
\delta_{mp}$, we write down the isospin interaction term as
\begin{equation}
  \label{eq:iso}
\mathcal{L}_{\mathrm{int}} \approx 2 g_{I} \psi_{\alpha n}^{\dagger}
\psi_{\alpha p} \psi_{\beta p}^{\dagger} \psi_{\beta n} + 4 g_{I}
\psi_{\alpha n}^{\dagger} \psi_{\alpha p} h^{\dagger}_{p} h_{n}\, ,
\end{equation}
where the local interactions of the holons are absorbed into the
usual $\varphi^{4}$ term called self-interactions.

The spinon-exchange interaction, i.e., the first term of
Eq.~(\ref{eq:iso}), can be ignored in the SM phase, while the electron
resonance term, which is given as the second one of
Eq.~(\ref{eq:iso}), will be allowed as quantum corrections
later. Then, it is straightforward to find an effective free energy
from the ``non-interacting'' theory
\begin{eqnarray}
  \label{eq:10}
F_{M}[\phi, \mu ; \delta, T ] &=& - \frac{N_{s}}{\beta} \sum_{k}
\Bigl\{ \ln \Bigl( 1 + e^{- \beta (E_{k}^{\psi} - \mu_{s} - i
\phi)} \Bigr) + \ln \Bigl( 1 + e^{- \beta ( E_{k}^{\psi} + \mu_{s}
+ i \phi)} \Bigr) \Bigr\} \cr && + \frac{1}{\beta} \sum_{q}
\Bigl\{ \ln \Bigl( 1 - e^{- \beta (E_{q}^{h} - \mu - i
\phi)}\Bigr) + \ln \Bigl( 1 - e^{- \beta ( E_{q}^{h} - \mu + i
\phi)}\Bigr) \Bigr\} + \mu \delta + \mu_{s}\, ,
\end{eqnarray}
where $E_{k}^{\psi} = k^{2}/2m_{\psi}$ and $E_{q}^{h} =
q^{2}/2m_{h}$ denote the dispersions for the spinons and holons,
respectively. The $N_{s}$ is the spin degeneracy, which is given by $N_{s}
= 2$ in the physical case. This effective free energy can be rewritten
in the following typical PNJL expression
\begin{eqnarray}
 F_{M}^{SM}[\Phi,\mu;\delta,T] &=&
-\frac{N_{s}}{\beta} \sum_{k} \ln \Bigl( 1 + 2 \bigl[ \Phi \cosh
\beta \mu_{s} - \sqrt{1-\Phi^{2}} \sinh \beta \mu_{s} \bigr] e^{-
\beta \frac{k^{2}}{2m_{\psi}} } + e^{- 2\beta
\frac{k^{2}}{2m_{\psi}} } \Bigr) \cr && + \frac{1}{\beta} \sum_{q}
\ln \Bigl( 1 - 2 \Phi e^{- \beta (\frac{q^{2}}{2m_{h}} - \mu )} +
e^{- 2 \beta (\frac{q^{2}}{2m_{h}} - \mu )} \Bigr) + \mu \delta +
\mu_{s}\, ,
\end{eqnarray}
where $\Phi = \cos \beta \phi$ is the Polyakov-loop parameter that is
gauge-invariant and physical. Minimizing the free energy with respect
to $\Phi$, one always finds $\Phi = 1$, so that Eq.~(\ref{eq:matter})
is reduced to a deconfined theory. Matter fluctuations favor the
deconfinement as expected.

The confinement of the spinons and holons can be realized by an
effective Polyakov-loop action from gauge dynamics. Although an
explicit form of the Yang-Mills action with the Polyakov-loop
order parameter is much complex, one can derive an effective
theory of the Polyakov-loop parameter in principle, integrating
over quantum fluctuations. Actually, performing $\int D
a_{\mu}^{1} D a_{\mu}^{2}$ in the one-loop level, one
finds~\cite{Weiss}
\begin{equation}
F_{G}[\Phi;T] \;=\; \frac{1}{2 \pi \beta} \int_{0}^{\infty} d k k
\ln (1 - 2 \Phi e^{-\beta k} + e^{-2 \beta k} )\, ,
\label{eq:free_energy}
\end{equation}
well shown in appendix A. Unfortunately, the free energy for the
gauge field from one-loop approximation always gives rise to $\Phi
= 1$ that corresponds to the deconfinement phase~\cite{Weiss}. It
is necessary to take quantum fluctuations into account in a
non-perturbative way. Though such a procedure is not theoretically
known yet, we will construct an effective free energy as follows
\begin{eqnarray}
F_{G}[\Phi;T] = A_{4} T^{3} \Bigl\{ \frac{A_{2} T_{0}}{A_{4}}
\Bigl( 1 - \frac{T_{0}}{T} \Bigr) \Phi^{2} - \frac{A_{3}}{A_{4}}
\Phi^{3} + \Phi^{4} \Bigr\}\, ,
\end{eqnarray}
where the constants $A_{i=2,3,4}$ are positive definite, and
$T_{0}$ is identified with the critical temperature for the
confinement-deconfinement transition (CDT). Since the CDT is known
as the first order from the lattice simulation~\cite{PNJL_Models},
the cubic-power term with a negative constant is introduced such
that $\Phi = 0$ in $T < T_{0}$ while $\Phi = 1$ in $T > T_{0}$,
corresponding to the center symmetry ($Z_{2}$)
breaking~\cite{Fukushima}.

Combining both matter and gauge sectors, we obtain an effective
PNJL free energy for the SM phase
\begin{eqnarray}
  \label{eq:pnjlfree}
F_{PNJL}[\Phi,\mu;\delta,T] &=& F_{M}[\Phi,\mu;\delta,T] +
F_{G}[\Phi;T]  \cr
&=& -\frac{N_{s}}{\beta} \sum_{k} \ln \Bigl( 1
+ 2 \bigl[ \Phi \cosh \beta \mu_{s} - \sqrt{1-\Phi^{2}} \sinh
\beta \mu_{s} \bigr] e^{- \beta \frac{k^{2}}{2m_{\psi}} } + e^{-
2\beta \frac{k^{2}}{2m_{\psi}} } \Bigr) \cr
&& +\; \frac{1}{\beta}
\sum_{q} \ln \Bigl( 1 - 2 \Phi e^{- \beta (\frac{q^{2}}{2m_{h}} -
\mu )} + e^{- 2 \beta (\frac{q^{2}}{2m_{h}} - \mu )} \Bigr) + \mu
\delta + \mu_{s} \cr && +\; A_{4} T^{3} \Bigl\{ \frac{A_{2}
T_{0}}{A_{4}} \Bigl( 1 - \frac{T_{0}}{T} \Bigr) \Phi^{2} -
\frac{A_{3}}{A_{4}} \Phi^{3} + \Phi^{4} \Bigr\} .
\end{eqnarray}
The CDT is driven by the gauge sector while the
matter fluctuations turn the first-order transition into the
confinement-deconfinement crossover (CDC) because the $Z_{2}$
center symmetry is explicitly broken in the presence of matters,
so that the Polyakov-loop does not become an order parameter in a
rigorous sense~\cite{Fukushima}. One may regard this PNJL
construction as our point of view for the present problem,
motivated from the crossover without the Higgs mechanism.
Actually, one can construct the PNJL free energy. Precisely
speaking, the gauge sector results in $\Phi = 0$ in $T < T_{CD}$
and $\Phi = 1$ in $T > T_{CD}$. Note that $T_{CD}$ is the CDC
temperature in the presence of matters, which is smaller than $T_{0}$
because matters favor the deconfinement.

\begin{figure}[ht]
\vspace{0.5cm}

\centerline{\includegraphics[scale=1.2]{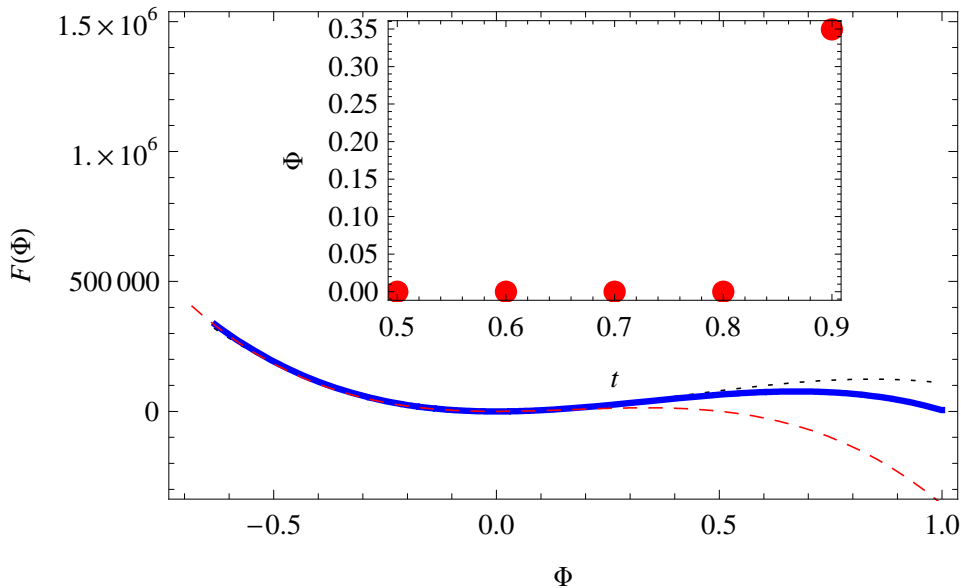} } \caption{
(Color online) The effective PNJL free energy as a function of the
Polyakov-loop parameter with $T < T_{CD}$ (Black-Dotted), $T =
T_{CD}$ (Blue-Thick), and $T > T_{CD}$ (Red-Dashed). Inset: The
Polyakov-loop parameter as a function of temperature scaled with
$T_{0}$. The Polyakov-loop parameter vanishes in $T < T_{CD}$,
causing confinement, while it becomes condensed in $T > T_{CD}$,
resulting in deconfinement.}
\label{fig:1} 
\end{figure}

Figure~\ref{fig:1} shows the free energy as a function of the
Polyakov-loop order parameter with temperature varied. The free energy
at $\Phi = 0$ in $T < T_{CD}$ is drawn in the dotted curve, whereas
that at $\Phi = 1$ in $T > T_{CD}$ is depicted in the dashed one. The
thick solid curve represents the case of $T = T_{CD}$. An interesting
point is that the holon chemical potential of a negative value is much
larger in the confinement phase than that in the deconfinement phase,
being consistent with confinement. The inset of Fig.~\ref{fig:1}
displays the Polyakov-loop parameter that starts to appear around $T =
T_{CD} < T_{0}$, where $T_{0}$ is the confinement-deconfinement
transition temperature of the pure gauge sector. 

Interestingly, the result in the mean-field approach of the PNJL model
implies that the condensation of holons is not allowed, since
$D[\Phi,\mu] = 1 - 2 \Phi \exp{[- \beta (q^{2}/2m_{h} - \mu
)]} + \exp{[- 2 \beta (q^{2}/2m_{h} - \mu )]}$ of the holon sector in
Eq.(\ref{eq:pnjlfree}) cannot reach the zero value
because of $0 \leq \Phi < 1$ except for $\Phi = 1$. In other
words, Higgs phenomena are not compatible with the confinement in
this description, being not inconsistent with the previous
field-theoretic result~\cite{Higgs_Confinement}. It should be also
noted that the mean-field approximation does not take into account
feedback effects from matters to gauge fluctuations. In fact,
The fluctuations of the Fermi surface are not introduced, so that
Landau-damped dynamics for gauge fluctuations is still missing. Thus, 
it is desirable to introduce quantum corrections beyond the
mean-field approximation in the PNJL, which is the main topic of the
work.

Since the Higgs phase is not allowed in the presence of the
Polyakov-loop order parameter, an immediate issue boils down to how to
describe the Fermi liquid phase. We will now examine the electron
self-energy in the confinement phase, certainly recovering the
Fermi liquid self-energy proportional to $\omega^{2}$ with
frequency $\omega$ below a certain temperature associated with the
holon chemical potential or the holon mass gap.

\section{Crossover from non-Fermi liquid to Fermi liquid}
The central question of the present study is about the fate of the
spinons and holons when the Polyakov-loop parameter vanishes. The
spinon-holon coupling term in Eq.~(\ref{eq:iso}) can be expressed as
follows
\begin{equation}
\mathcal{S}_{\mathrm{el}} \;=\; \int_{0}^{\beta} d \tau \int d^{2} r \left(
\psi_{\sigma n}^{\dagger} h_{n} c_{\sigma} + c_{\sigma}^{\dagger}
h^{\dagger}_{p} \psi_{\sigma p} - \frac{1}{g_{I}}
c_{\sigma}^{\dagger} c_{\sigma} \right)  ,
\end{equation}
where $\sigma$ and $n(p)$ represent spin and SU(2) indices,
respectively. Since the Grassmann variable $c_{\sigma}$ carries
exactly the same quantum numbers as the electron, one may identify
it as the Hubbard-Stratonovich field $c_{\sigma}$. The effective
coupling constant $g_{I}$ plays a role of the chemical potential
for the electrons. Note that the Fermi surface of the electrons
differs from that of the spinons in principle.

One can introduce quantum corrections self-consistently in the
Luttinger-Ward functional approach~\cite{LW}, in which only
self-energy diagrams are taken into account, vertex
corrections~\cite{Kim_LW} being ignored,
\begin{eqnarray}
  \label{eq:13}
F_{LW}[\phi,
\mu ; \delta, T ] &=& - \frac{1}{\beta} \sum_{i\omega} \sum_{k}
\Bigl\{ \ln \Bigl( - G^{\psi -1}_{\sigma\sigma,pp'} (k,i\omega)
\Bigr) + \Sigma^{\psi}_{\sigma\sigma,pp'}(k,i\omega)
G^{\psi}_{\sigma\sigma,p'p} (k,i\omega) \Bigr\} \cr && -\;
\frac{1}{\beta} \sum_{i\omega} \sum_{k} \Bigl\{ \ln \Bigl( - G^{c
-1}_{\sigma\sigma} (k,i\omega) \Bigr) +
\Sigma^{c}_{\sigma\sigma}(k,i\omega) G^{c}_{\sigma\sigma}
(k,i\omega) \Bigr\} \cr && +\; \frac{1}{\beta} \sum_{i\Omega}
\sum_{q} \Bigl\{ \ln \Bigl( - G^{h -1}_{pp'}(q,i\Omega) \Bigr) +
\Sigma^{h}_{pp'}(q,i\Omega) G^{h}_{p'p}(q,i\Omega) \Bigr\} \cr &&
-\; \frac{1}{\beta} \sum_{i\omega} \sum_{k} \frac{1}{\beta}
\sum_{i\Omega} \sum_{q} G^{c}_{\sigma\sigma} (k+q,i\omega+i\Omega)
G^{h}_{p'p}(q,i\Omega) G^{\psi}_{\sigma\sigma,pp'} (k,i\omega) +
\mu \delta + \mu_{s}\, .
\end{eqnarray}
Here, $G^{c}_{\sigma\sigma}
(k,i\omega)$, $G^{\psi}_{\sigma\sigma,pp'} (k,i\omega)$, and
$G^{h}_{pp'}(q,i\Omega)$ represent the single-particle Green functions
for confined electrons, spinons, and holons, respectively, written as
\begin{eqnarray}
  \label{eq:14}
- G^{c -1}_{\sigma\sigma} (k,i\omega) &=&
\Sigma^{c}_{\sigma\sigma}(k,i\omega) - g_{I}^{-1} ,\cr
 -G^{\psi -1}_{\sigma\sigma,pp'} (k,i\omega) &=& [- i (\omega + p
\phi) - p \mu_{s}] \delta_{pp'} + \frac{k^{2}}{2m_{\psi}}
\delta_{pp'} + \Sigma^{\psi}_{\sigma\sigma,pp'} (k,i\omega) , \cr
- G^{h -1}_{pp'}(q,i\Omega) &=& [- i ( \Omega + p \phi) - \mu]
\delta_{pp'} + \frac{q^{2}}{2m_{h}} \delta_{pp'} +
\Sigma^{h}_{pp'}(q,i\Omega) ,
\end{eqnarray}
where $\Sigma^{c}_{\sigma\sigma}(k,i\omega)$,
$\Sigma^{\psi}_{\sigma\sigma,pp'} (k,i\omega)$, and
$\Sigma^{h}_{pp'}(q,i\Omega)$ designate the self-energy
corrections for confined electrons, spinons, and holons,
respectively. This approximation is referred to as the Eliashberg
theory that was previously believed to be valid in the large-$N$
limit~\cite{FMQCP}, where $N$ represents the spin degeneracy.
However, the validity of the large-$N$ limit has been
questioned~\cite{SungSik_Genus,Metlitski_Sachdev1,Metlitski_Sachdev2}
recently, which will be discussed in the final section.

Having minimized the free energy with respect to each self-energy 
\begin{equation}
  \label{eq:15}
 \frac{\partial
F_{LW}(x,T)}{\partial \Sigma^{c}_{\sigma\sigma}(k,i\omega)} = 0 ,
~~~~~ \frac{\partial F_{LW}(x,T)}{\partial
\Sigma^{\psi}_{\sigma\sigma,pp'} (k,i\omega)} = 0 , ~~~~~
\frac{\partial F_{LW}(x,T)}{\partial \Sigma^{h}_{pp'}(q,i\Omega)}
= 0\, ,
\end{equation}
we find the self-consistent equations for the self-energies as follows
\begin{eqnarray}
  \label{eq:16}
\Sigma^{c}_{\sigma\sigma}(k,i\omega) &=& -
\frac{1}{\beta} \sum_{i\Omega} \sum_{q} G^{h}_{p'p}(q,i\Omega)
G^{\psi}_{\sigma\sigma,pp'} (k-q,i\omega-i\Omega) , \cr
\Sigma^{\psi}_{\sigma\sigma,pp'} (k,i\omega) &=& - \frac{1}{\beta}
\sum_{i\Omega} \sum_{q} G^{c}_{\sigma\sigma} (k+q,i\omega+i\Omega)
G^{h}_{p'p}(q,i\Omega) , \cr  \Sigma^{h}_{pp'}(q,i\Omega) &=&
\frac{1}{\beta} \sum_{i\omega} \sum_{k} G^{c}_{\sigma\sigma}
(k+q,i\omega+i\Omega) G^{\psi}_{\sigma\sigma,pp'} (k,i\omega)\, .
\end{eqnarray}
These equations were intensively discussed in
the context of heavy fermions \cite{Pepin_Paul} without the
confinement, i.e., without the Polyakov-loop parameter. Thus, this
framework may be regarded as an extension of the mean-field
analysis, both the confinement effects and the quantum
fluctuations being incorporated self-consistently.

\begin{figure}[htp]
\centerline{\includegraphics[scale=0.45]{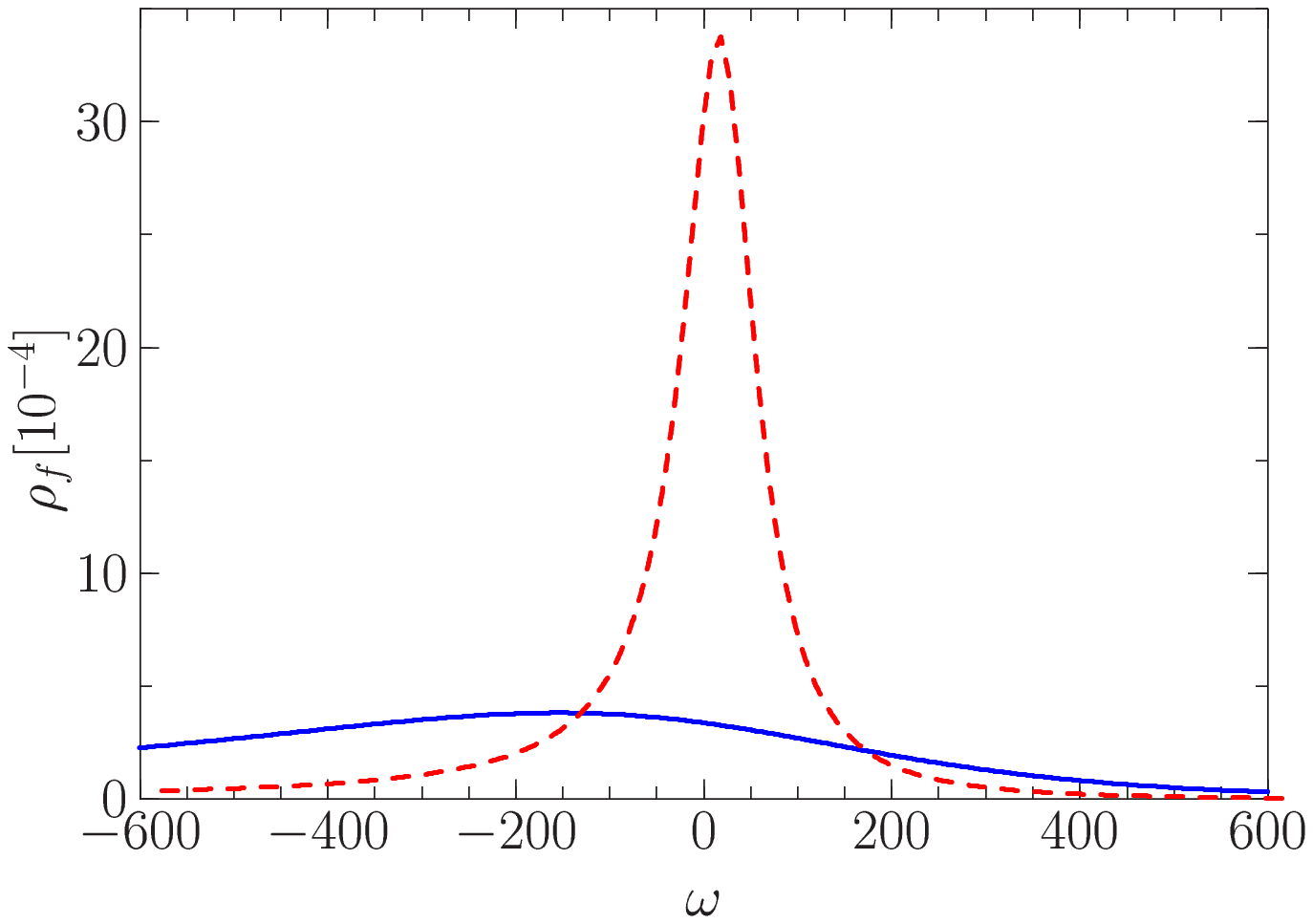}
\includegraphics[scale=0.45]{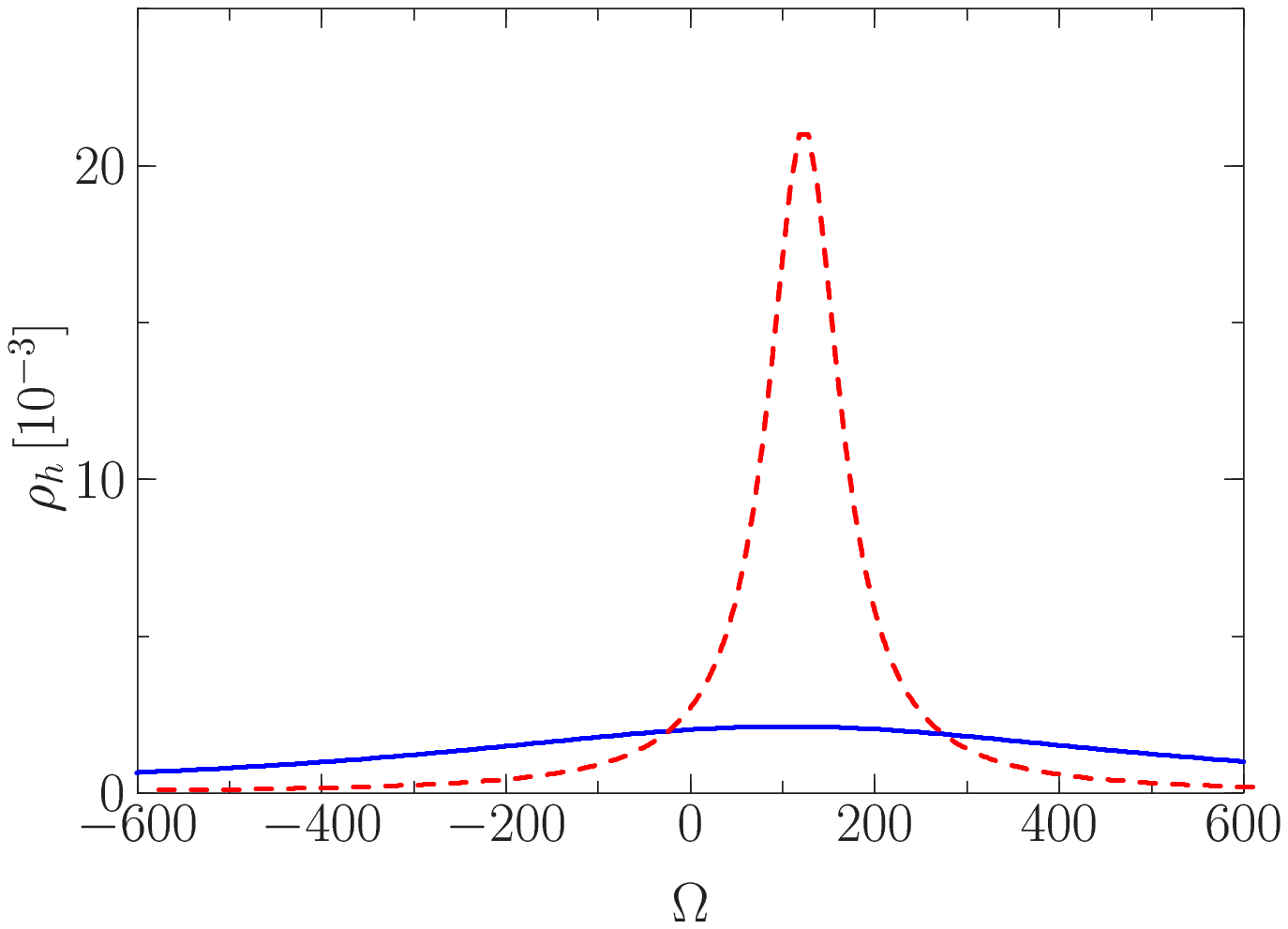}} \vspace{0.5cm}

\centerline{\includegraphics[scale=0.8]{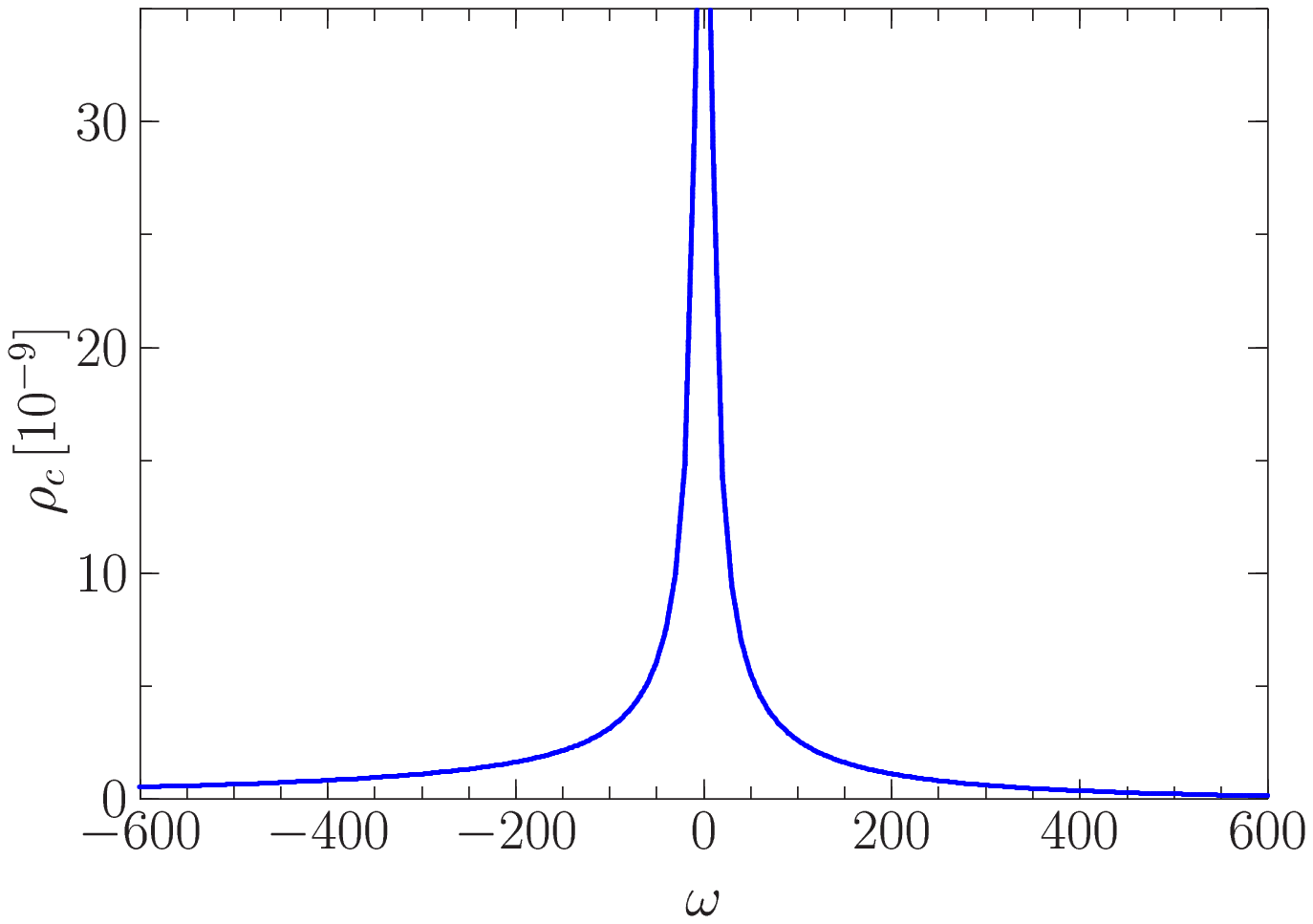}}
\caption{ (Color online) Spinon, holon, and electron spectra: The
Polyakov-loop parameter gives rise to huge imaginary parts for
self-energies of both spinons and holons in the confinement
region, making the spinon and holon spectra broader (blue thick
curve). On the other hand, electron excitations become rather
sharply defined due to confinement, but not fully coherent at high
temperatures, which originates from strong inelastic scattering
with such incoherent spinon and holon excitations.} \label{fig2}
\end{figure}

In the confinement phase the spectral function of the spinon
should not be reduced to the delta function owing to the presence
of the background potential $\phi$ even if the self-energy
correction is ignored. Actually, the Polyakov-loop parameter plays
a role of the imaginary part of the self-energy, which makes the
spinon resonance disappear as shown in the first panel of
Fig.~\ref{fig2}. The holon spectrum also features a broad
structure, presented in the second panel of Fig.~\ref{fig2}. It
indicates that both the spinon and the holon are not well-defined
excitations in the confinement phase. On the other hand, the
electron as a spinon-holon composite exhibits a rather sharp peak
in the last graph of Fig.~\ref{fig2}, since the imaginary part of
their self-energy vanishes at the Fermi surface in spite of no
pole structure in the electron Green function.

The holon self-energy is found to be of the standard form in two
dimensions:
\begin{eqnarray}
\Sigma_{p}^{b}(q,i\Omega) - \Sigma_{p}^{b}(q,0) &=&
- \frac{\rho_{c} }{i(\alpha-1)}
\Bigl\{\tan^{-1}\Bigl(\frac{i\Omega+ip\phi- v_{F}^{c} q^{*} +
v_{F}^{c} q}{- i\Omega - i p\phi + v_{F}^{c} q^{*} + v_{F}^{c}
q}\Bigr) \cr &&
\hspace{1.5cm} - \tan^{-1}\Bigl(\frac{i\Omega+ip\phi-\alpha v_{F}^{c}
q^{*} + \alpha v_{F}^{c} q}{- i\Omega - i p\phi + \alpha v_{F}^{c}
q^{*} + \alpha v_{F}^{c} q}\Bigr) \Bigr\}
\label{eq:Green}
\end{eqnarray}
except for $i\Omega \rightarrow i\Omega+ip\phi$. The $\rho_{c}$
denotes the density of states for the confined electron, and
$v_{F}^{c}$ stands for the corresponding Fermi velocity. The
$\alpha$ represents the ratio of the electron band mass to the
spinon one, given as almost unity. The $q^{*}$ designates the
Fermi-momentum mismatch between the confined electron and the
spinon. A detailed derivation is shown in Appendix B.

Inserting Eq.~(\ref{eq:Green}) into the electron self-energy
equation, we can find its explicit form. See Appendix C for its
derivation. An important energy scale is given by the holon
chemical potential $\mu$. In $T > |\mu|$, holon dynamics is
described by the dynamical exponent $z = 3$, resulting from the
Landau damping of the electron and
spinon~\cite{Kim_LW,Pepin_Paul}. The imaginary part of the
self-energy turns out to be proportional to $T^{2/3}$, since the
confined electrons are scattered with such $z = 3$ dissipative
modes~\cite{Kim_LW,Pepin_Paul}. On the other hand, the holon
excitations have gaps in $T < |\mu|$, so that scattering with
confined electrons are suppressed, which recovers the Fermi
liquid. Thus, the Fermi liquid appears as the coherence effect in
the confinement phase rather than the Higgs phase in the
deconfinement state. This mechanism resolves the artificial
transition at finite temperatures.

\begin{figure}[htp]
\centerline{\includegraphics[scale=0.8]{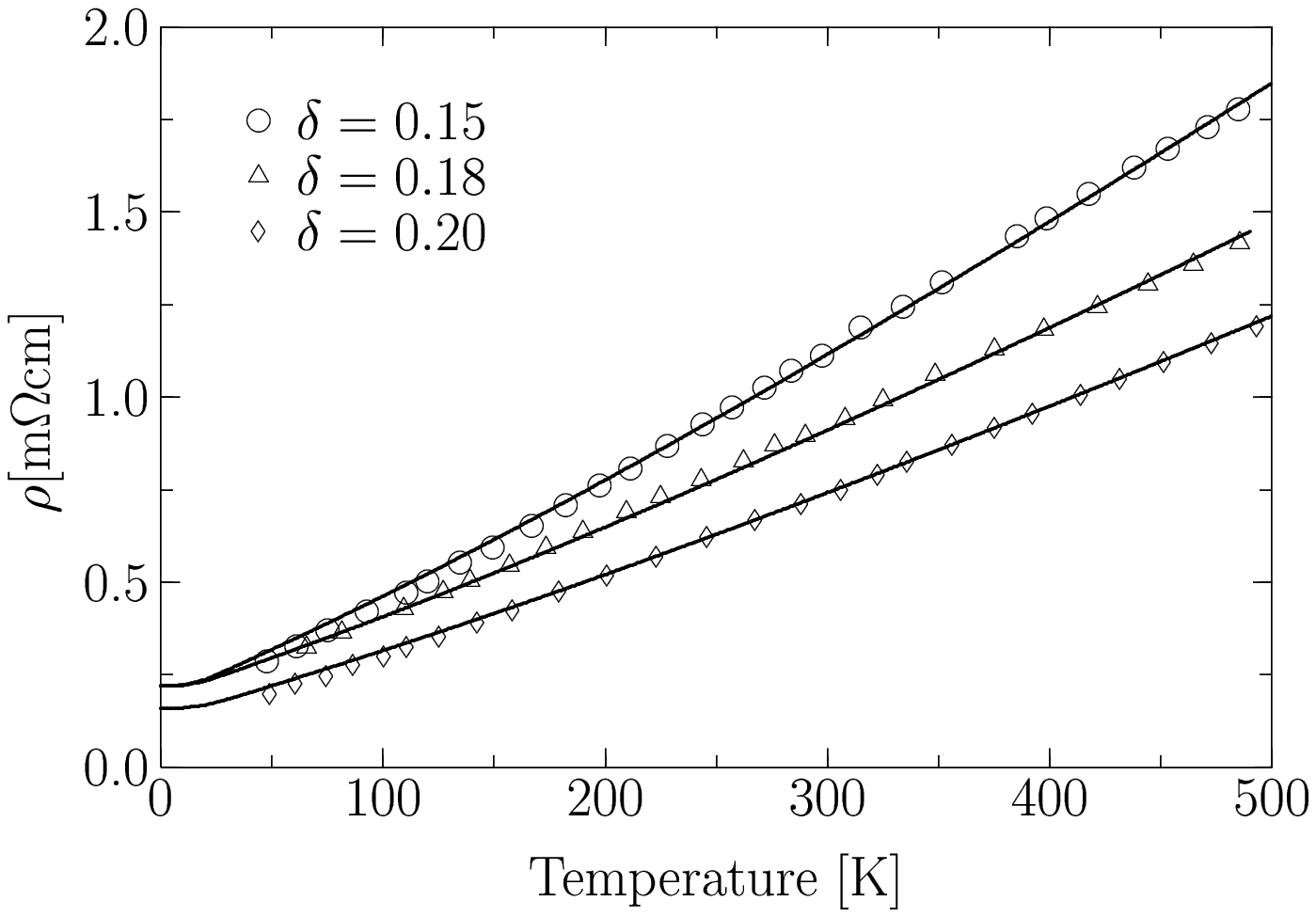}} \caption{
(Color online) The results of the electrical resistivity are drawn
as a function of temperature. The circle, triangle, and lozenge
denote the experimental data~\cite{Data} corresponding to
different values of the doping. The non-Fermi liquid electrical
resistivity \cite{Data_fitting} evolves into the typical
Fermi-liquid behavior below the crossover temperature, given by
the holon gap.} \label{fig3}
\end{figure}

The coherence crossover can be manifestly observed in the
electrical transport. It should be realized that the Ioffe-Larkin
composition rule for the transport~\cite{IoffeLarkin} does not
apply to the confinement phase. Instead, the electrical currents
would be carried by confined electrons dominantly. This is in
sharp contrast with the existing gauge-theoretic description for
the SM phase, where the anomalous electrical transport originates
from holon dynamics interacting with gauge
fluctuations~\cite{SM_U1GT}. In addition to this important aspect,
the relaxation time associated with the imaginary part of the
self-energy differs from the transport time, and the back
scattering contribution is factored out by the vertex corrections.
Actually, the role of the vertex corrections has been investigated
carefully in the context of the heavy fermion quantum criticality
of the Kondo breakdown~\cite{Pepin_Paul,Kim_TR,Kim_QBE_KB}.
Employing the quantum Boltzmann equation approach,
Refs.~\cite{Pepin_Paul,Kim_TR,Kim_QBE_KB} have shown that the
backscattering contribution selected from the vertex corrections
is given by $1 - \alpha \cos \theta$ in the scattering mechanism
from spinon-holon binding fluctuations, where $\theta$ is an angle
measured from the direction of the electric current. An important
point is that this scattering mechanism introduces the factor
$\alpha$, which is negligibly small in the Kondo breakdown
scenario, because the band mass of the spinons is large due to the
$f$-orbital character. On the other hand, we should have $\alpha
\approx 1$ in the SM phase because the band mass of the spinons is
almost identical with that of the emergent electrons. In this case
the vertex corrections should be introduced. Their temperature
dependence is well known to be proportional to $T^{2/3}$ for two
dimensional $z = 3$ fluctuations~\cite{Kim_TR}.

Introducing both the self-energy and vertex corrections, we reach the
final expression of the electrical resistivity
\begin{equation}
\rho_{el}(T) \;=\; \rho_{0} + \mathcal{C} \Bigl( N_{s} \rho_{c}
\frac{v_{F}^{c2}}{3} \Bigr)^{-1} T^{2/3} Im[\Sigma_{c}(T)]\, ,
\end{equation}
where $\rho_{0}$, $\mathcal{C}$, and $N_{s}$ denote, respectively,
the residual resistivity due to disorder, the strength for the vertex
corrections, and the spin degeneracy, among which $\rho_0$ and
$\mathcal{C}$ are free parameters. Note that the $T^{2/3}$ factor
results from the vertex corrections. As shown in Fig.~\ref{fig3}, the
results are in remarkable agreement with the experimental data, which
supports our confinement scenario. In addition, the $T^{2}$ behavior
is clearly observed at low temperatures, confirming our statement that
the crossover from the SM phase to the Fermi liquid state is described
by the coherence effect with the confinement.

\section{Conclusion : Loose end}

In the present study we proposed a novel mechanism for the
marginal Fermi-liquid phenomenology near optimal doping in high
$T_{c}$ cuprates. First, we tried to resolve the artificial
finite-temperature second-order transition from a spinon-holon
deconfined non-Fermi liquid state (or a finite-temperature
spin-liquid state) to a spinon-holon Higgs-confined Fermi-liquid
phase. Indeed, we made the continuous transition smoothen into the
crossover, but from a spinon-holon confined incoherent
Fermi-liquid state to a spinon-holon confined coherent
Fermi-liquid phase. The confinement of spinons and holons at
finite temperatures was phenomenologically described by the
Polyakov-loop parameter, where confined spinon-holon composite
particles are identified with incoherent electrons in the
temperature-regime above the "holon" chemical potential.
Although holons are not elementary excitations due to confinement,
given by the huge imaginary part from the Polyakov-loop parameter,
their chemical potential defines the coherence energy scale.

The second point, which is more important in our opinion, is that we 
proposed one possible mechanism for the non-Fermi liquid
resistivity near the optimal doping, introducing the Polyakov-loop
parameter. This object gives rise to not only confinement but also
an interesting self-energy correction to such confined spinons,
where the imaginary part of the spinon self-energy is proportional
to temperature, regarded as the signature of confinement. This
anomalous dynamics affects dynamics of incoherent electrons,
responsible for the non-Fermi liquid resistivity above the
Fermi-liquid coherence temperature near the optimal doping. We
believe that this is a novel mechanism for the non-Fermi liquid
transport around the optimal doping in high $T_c$ cuprates. In
order to realize such non-Fermi liquid physics, we usually resort
to some type of critical fluctuations, which can give rise to the
T-linear self-energy correction. Recall the marginal Fermi-liquid
Ansatz. In this study we do not assume any kinds of symmetry
breaking for the scattering mechanism. The Polyakov-loop parameter
serves a novel scattering mechanism for the marginal Fermi-liquid
physics near the optimal doping.

In fact, we cannot justify the fundamental reason why the
Polyakov-loop parameter should be utilized near optimal doping.
Our formulation is based on a spin liquid state, expected to apply
to an underdoped paramagnetic insulating-like state (at least
applicable to finite temperatures, usually referred to as the
proximity effect of the spin liquid physics), where spinons and
holons appear as elementary excitations. Although we cannot
explain how such deconfined particles evolve into confined
electrons as a function of hole concentration, we believe that
these deconfined excitations should be confined around the optimal
doping region above the pseudogap phase, where the anomalous
$T$-linear-like resistivity is shown. An important question is how
to construct incoherent electrons from spinons and holons at
finite temperatures. One can introduce a spinon-holon bound state
based on the ladder-type t-matrix approximation without the
Polyakov-loop parameter. However, it is not clear whether this
construction allows the non-Fermi liquid resistivity or not. We
would like to emphasize again that the Polyakov-loop parameter can
solve two problems at the same time, confinement (conceptually
valid) and non-Fermi liquid (marginal Fermi liquid) physics near
the optimal doping.

If one insists that deconfinement should be stabilized due to
the Fermi surface \cite{SungSik_Deconfinement} and the confined
Fermi liquid state results from the Anderson-Higgs mechanism, we
would like to point out that the existence argument can be
justified in the large-$N$ limit, where $N$ is the number of
degeneracy in fermion flavors. If we consider a realistic
situation at finite $N$, several types of missing correlations may
play an important role for fermion dynamics, where enhanced
correlations will reduce conductivity, which means that such
fermions can have difficulty in screening gauge interactions. It
is natural to expect that fermions in an almost insulating state
will not screen gauge forces appropriately, even if they form
their Fermi surface \cite{Kim_FS_Deconfinement}. Although we
cannot give any definite statement on confinement in the presence
of the Fermi surface, we believe that there is no consensus, in
particular, when finite $N$ should be considered
\cite{Finite_N_Confinement}. What we can argue here is that our
confinement formulation based on the Polyakov-loop parameter turns
the artificial second-order transition into the crossover,
conceptually more advantageous than the holon condensation.

One may suspect whether our Fermi liquid state is the genuine
Fermi liquid phase or not because it is not obvious that the Fermi
surface of such coherent electrons satisfies the Luttinger theorem
\cite{Luttinger_Theorem}. The main quantity is the chemical
potential of electrons, emerging from confinement of spinons and
holons. Although this quantity should be and can be determined
self-consistently in our formulation, we did not perform such a
serious analysis. That analysis will involve many delicate issues,
for example, a more accurate expression of the self-energy
correction of electrons in order to determine the chemical
potential accurately. In our study we just introduce the electron
chemical potential phenomenologically, which can be determined to
satisfy the Luttinger theorem. This is the best at present in our
formulation, which is another reason why we call our theory a
phenomenological description.

We would like to point out that the issue on the formation of
incoherent electrons from deconfined spinons and holons is not
seriously discussed in our study. Incoherent electrons are assumed
to exist at finite temperatures around the optimal doping region,
resorting to the Polyakov-loop parameter. Of course, we pointed
out that spinons and holons can be deconfined at high
temperatures, showing that the Polyakob-loop parameter is
non-vanishing in the saddle-point approximation. In our mean-field
analysis the change from the spinon-holon deconfined state to the
confined incoherent Fermi-liquid state (by Polyakov loop) is
naturally interpreted as the crossover at finite temperatures
(much higher than the non-Fermi liquid to Fermi liquid crossover
temperature) because there is no symmetry breaking, as discussed
before. Unfortunately, it doesn't seem to be clear whether it is
crossover or not beyond the mean-field analysis, which means to
introduce quantum fluctuations for the Polyakov-loop parameter
\cite{Deconfinement_KT}. It is also an open question in our
description how the Polyakov-loop parameter evolves by hole
doping. We believe that this requires some modifications of our
present formulation.

One may criticize cautiously the Eliashberg approximation for
neglecting the vertex corrections for self-energies. It has been
recently clarified that two dimensional Fermi surface is still
strongly interacting even in the large-$N$
limit~\cite{SungSik_Genus,Metlitski_Sachdev1,Metlitski_Sachdev2,
McGreevy,Kim_Ladder,Kim_AL}, which implies that vertex corrections
should be incorporated. An important question is raised as to
whether these vertex corrections give rise to novel critical
exponents beyond the Eliashberg theory. Several perturbative
analysis demonstrated that although ladder-type vertex corrections
do not change the critical exponents of the Eliashberg theory,
Aslamasov-Larkin corrections bring about modifications for such
critical exponents~\cite{Metlitski_Sachdev1,Metlitski_Sachdev2}.
If this is a general feature beyond this level of approximation,
various quantum critical phenomena~\cite{HFQCP} should be
reconsidered because the novel anomalous exponents in the fermion
self-energy corrections are expected to affect various novel
critical exponents for thermodynamics, transport, etc. However,
considering our recent experiences in comparison with various
experiments in heavy fermion quantum criticality, we want to
emphasize that the critical exponents based on the Eliashberg
theory explain thermodynamics \cite{Kim_Adel_Pepin}, both
electrical and thermal transport coefficients \cite{Kim_TR},
uniform spin susceptibility \cite{Kim_Jia}, and Seebeck effect
\cite{Kim_Pepin_Seebeck} noticeably well.

Recently, one of the authors investigated the role of the vertex
corrections non-perturbatively, summing them to
infinite order~\cite{Kim_Ladder,Kim_AL}. It turns out that
particular vertex corrections given by ladder diagrams do not
change Eliashberg critical exponents at all \cite{Kim_Ladder},
being consistent with the perturbative analysis. This was
performed in a fully self-consistent way, the Ward identity being
utilized. In contrast with the previous perturbative analysis, the
Aslamasov-Larkin corrections were shown not to modify the
Eliashberg dynamics~\cite{Kim_AL}. In analogy with
superconductivity, where the superconducting instability described
by the Aslamasov-Larkin vertex corrections is reformulated by the
anomalous self-energy in the Eliashberg framework of the Nambu
spinor representation \cite{BCS}, we claimed that the off-diagonal
self-energy associated with the 2$k_{F}$ particle-hole channel
incorporates the same (Aslamasov-Larkin) class of quantum
corrections in the Nambu spinor representation. We evaluated an
anomalous pairing self-energy in the Nambu-Eliashberg
approximation, which vanishes at zero energy but displays the same
power-law dependence for the frequency as the normal Eliashberg
self-energy. As a result, even the pairing self-energy corrections
do not modify the Eliashberg dynamics without the Nambu spinor
representation. We leave this profound and important issue for future
investigations.

It will be of great interest to apply the PNJL scheme to the
spin-liquid theory \cite{Spin_Liquid} and Kondo breakdown scenario
\cite{Pepin_Paul} for heavy fermions. In particular, the
artificial finite-temperature transition would be resolved for the
heavy fermion transition in the same way as the present case,
which may allow a new scenario of quantum criticality.

\section*{Acknowledgments}
The authors are grateful to K. Fukushima for helpful comments.
K.-S. Kim was supported by the National Research Foundation of
Korea (NRF) grant funded by the Korea government (MEST) (Grant No.
2011-0074542). The present work (H.-Ch. Kim) is also supported by
Basic Science Research Program through the NRF funded by the MEST
(Grant No. 2010-0016265).

\appendix

\section{An effective Polyakov-loop action in the one loop level from 
  the Yang-Mills theory} 

Taking the reparameterization 
\begin{equation}
a_{\mu}^{+} \;=\;
\frac{a_{\mu}^{1} + a_{\mu}^{2}}{\sqrt{2}} , ~~~~~ a_{\mu}^{-} =
\frac{a_{\mu}^{+} - a_{\mu}^{2}}{\sqrt{2} i}\, , 
\end{equation}
one can rewrite the Yang-Mills action as follows 
\begin{eqnarray}
\mathcal{L}_{YM} &=&
\frac{1}{4g^{2}}\Bigl( [\partial_{\mu} a_{\nu}^{+} -
\partial_{\nu} a_{\mu}^{+} - i g (a_{\mu}^{3} a_{\nu}^{+} -
a_{\nu}^{3} a_{\mu}^{+})]^{2} + [\partial_{\mu} a_{\nu}^{-} -
\partial_{\nu} a_{\mu}^{-} + i g (a_{\mu}^{3} a_{\nu}^{-} -
a_{\nu}^{3} a_{\mu}^{-})]^{2} \cr 
& & +\; [\partial_{\mu} a_{\nu}^{3}
- \partial_{\nu} a_{\mu}^{3} - i g (a_{\mu}^{+} a_{\nu}^{-} -
a_{\nu}^{+} a_{\mu}^{-})]^{2}  \Bigr) \cr 
& = & \frac{1}{4g^{2}}
\Bigl( [(\partial_{\tau} - i g a_{\tau}^{3})a_{i}^{+} -
(\partial_{i} - i g a_{i}^{3})a_{\tau}^{+} ]^{2} + (i
\longleftrightarrow \tau) + [(\partial_{i} - i g
a_{i}^{3})a_{j}^{+} - (\partial_{j} - i g a_{j}^{3})a_{i}^{+}
]^{2} \Bigr) \cr && +\; \frac{1}{4g^{2}} \Bigl( [(\partial_{\tau} +
i g a_{\tau}^{3})a_{i}^{-} - (\partial_{i} + i g
a_{i}^{3})a_{\tau}^{-} ]^{2} + (i \longleftrightarrow \tau) +
[(\partial_{i} + i g a_{i}^{3})a_{j}^{-} - (\partial_{j} + i g
a_{j}^{3})a_{i}^{-} ]^{2} \Bigr) \cr && +\; \frac{1}{4g^{2}} \Bigl(
[\partial_{\tau} a_{i}^{3} - \partial_{i} a_{\tau}^{3} - i g
(a_{\tau}^{+} a_{i}^{-} - a_{i}^{+} a_{\tau}^{-})]^{2} + (i
\longleftrightarrow \tau) + [\partial_{i} a_{j}^{3} - \partial_{j}
a_{i}^{3} - i g (a_{i}^{+} a_{j}^{-} - a_{j}^{+} a_{i}^{-})]^{2}
\Bigr)\, , 
\end{eqnarray}
where $a_{\mu}^{3}$ is introduced as the covariant
derivative for the $a_{\mu}^{\pm}$ field.

Taking the Ansatz 
\begin{equation}
a_{\tau}^{3} = \phi, \;\;\;\; a_{i}^{3} = 0  
\end{equation}
for the lowest-order approximation, we reach the following
expression for the Yang-Mills action 
\begin{eqnarray}
{\cal L}_{YM} &\approx&
\frac{1}{2g^{2}} \Bigl( [(\partial_{\tau} - i g \phi)a_{i}^{+} -
\partial_{i} a_{\tau}^{+} ]^{2} + [ \partial_{i} a_{j}^{+} -
\partial_{j} a_{i}^{+} ]^{2} \Bigr) + \frac{1}{4g^{2}} \Bigl(
[(\partial_{\tau} + i g \phi)a_{i}^{-} - \partial_{i} a_{\tau}^{-}
]^{2} + [ \partial_{i} a_{j}^{-} - \partial_{j} a_{i}^{-} ]^{2}
\Bigr) \cr 
&& +\; \frac{1}{2} [a_{\tau}^{+} a_{i}^{-} - a_{i}^{+}
a_{\tau}^{-}]^{2} + \frac{1}{4} [ a_{i}^{+} a_{j}^{-} - a_{j}^{+}
a_{i}^{-} ]^{2}\,. 
\end{eqnarray}
Integrating over $a_{\mu}^{\pm}$ without
their interactions given by the last two terms, we obtain an
effective action for the Polyakov-loop parameter in the one loop
level, corresponding to Eq.~(\ref{eq:free_energy}).

\section{Holon self-energy with the Polyakov-loop parameter}

Linearizing the band dispersion near each Fermi surface, we obtain
the electron and spinon Green functions
\begin{eqnarray}
G_{c}(k,i\omega)
&\approx& \frac{1}{i\omega - v_{F}^{c} k - \Sigma_{c}(i\omega)} ,
\cr  G_{\psi}(k+q,i\omega+i\Omega) &=& \frac{1}{ (i\omega +
i\Omega + i p \phi) - v_{F}^{\psi} k - v_{F}^{\psi} q_{\parallel}
- v_{F}^{\psi} q^{*} - \Sigma^{\psi}_{p} (i\omega+i\Omega)}\, ,
\end{eqnarray}
respectively, where each self-energy correction is assumed to
depend only on the frequency, which is well adopted in the
Eliashberg framework~\cite{FMQCP} because most singular
corrections result from the frequency dependence. In the electron
Green function we set $\Sigma_{c}(k,i\omega) \approx
\frac{k^{2}}{2m_{c}} + \Sigma_{c}(i\omega)$, and expand the
dispersion near the chemical potential $g^{-1}_{I}$. The Fermi
velocity $v_{F}^{\psi}$ of the spinons is related with that of the
electrons as $v_{F}^{\psi} = \alpha v_{F}^{c}$, where $\alpha
\approx 1$ because the band mass of the electrons is almost
identical with that of the spinons. $q^{*} = k_{F}^{c} -
k_{F}^{\psi}$ is the Fermi surface mismatch.

Since fermion self-energies are assumed to depend on frequency
only, the boson self-energy is basically the same as that in the
random-phase approximation~\cite{FMQCP}. Inserting the above
fermion Green functions into the self-consistent equation for
the boson self-energy, we obtain
\begin{eqnarray}
&& \Sigma_{p}^{b}(q,i\Omega)
- \Sigma_{p}^{b}(q,0)\cr
 &\approx& - \frac{ \rho_{c} }{2\pi^{2}}
\int_{-\infty}^{\infty} {d\omega} \int_{0}^{2\pi} {d \theta}
\int_{-\infty}^{\infty} d\epsilon \frac{1}{i\omega - \epsilon }
\frac{1}{(i\omega + i\Omega + i p \phi) - \alpha \epsilon - \alpha
v_{F}^{c} q^{*} - \alpha v_{F}^{c} q \cos\theta } \cr &=& i
\frac{ \rho_{c} }{2\pi } \int_{-\infty}^{\infty} {d\omega}
\int_{0}^{2\pi} {d \theta} \frac{\Theta(\omega + \Omega + p \phi)
- \Theta(\omega)}{i\Omega + (1-\alpha)i\omega + i p \phi - \alpha
v_{F}^{c} q^{*} - \alpha v_{F}^{c} q \cos\theta} \cr  &=& \frac{
\rho_{c} }{2} \int_{-\Omega-p\phi}^{0} {d\omega} \Bigl\{
[i\Omega+(1-\alpha)i\omega+ip\phi-\alpha v_{F}^{c} q^{*} + \alpha
v_{F}^{c} q ] [- i \Omega - (1-\alpha)i\omega - ip\phi +\alpha
v_{F}^{c} q^{*} + \alpha v_{F}^{c} q ] \Bigr\}^{-\frac{1}{2}}
\cr &=& - \frac{ \rho_{c} }{i(\alpha-1)}
\Bigl\{\tan^{-1}\Bigl(\frac{i\Omega+ip\phi- v_{F}^{c} q^{*} +
v_{F}^{c} q}{- i\Omega - i p\phi + v_{F}^{c} q^{*} + v_{F}^{c}
q}\Bigr) - \tan^{-1}\Bigl(\frac{i\Omega+ip\phi-\alpha v_{F}^{c}
q^{*} + \alpha v_{F}^{c} q}{- i\Omega - i p\phi + \alpha v_{F}^{c}
q^{*} + \alpha v_{F}^{c} q}\Bigr) \Bigr\}\, ,
\end{eqnarray}
where $\rho_{c}$ is the density of states for electrons. The
$\Sigma_{p}^{b}(q,0)$ is proportional to $q^{2}$.

\section{Electron self-energy with the Polyakov-loop parameter}
Since the boson self-energy is nothing but the typical expression
in the two dimensional Fermi liquid except for the Polyakov-loop
parameter, the boson self-energy is given by the Landau-damping,
that is, not by $\Omega$ alone but by $\Omega + p \phi$ in the
low-energy limit. Inserting this approximate boson self-energy with
the damping coefficient $\gamma_{b}$ into the electron self-energy
equation, we obtain
\begin{eqnarray}
&& \Sigma^{c}(k_{F}^{c},i\omega) \cr
& = &
- \frac{1}{\beta} \sum_{i\Omega} \int_{-\infty}^{\infty} \frac{d
q_{\parallel}}{2\pi} \int_{-\infty}^{\infty} \frac{d
q_{\perp}}{2\pi} \frac{1}{[- i ( \Omega + p \phi) - \mu] +
\frac{q_{\perp}^{2}}{2m_{h}} + \gamma_{b} \frac{|\Omega + p
\phi|}{q_{\perp}}}\cr
&&\hspace{3.8cm}\times  \frac{1}{- i (\omega - \Omega + p \phi) +
v_{F}^{\psi} q_{\parallel} + \Sigma^{\psi}_{p}
(k_{F}^{c}-q,i\omega-i\Omega)} \cr
& \approx& - \frac{1}{\beta}
\sum_{i\Omega} \int_{-\infty}^{\infty} \frac{d
q_{\parallel}}{2\pi} \frac{1}{- i (\omega - \Omega + p \phi) +
v_{F}^{\psi} q_{\parallel} } \frac{1}{[- i ( \Omega + p \phi) -
\mu] + \frac{q_{\perp}^{2}}{2m_{h}} + \gamma_{b} \frac{|\Omega + p
\phi|}{q_{\perp}}}\,  .
\end{eqnarray}

The spectral representation being employed, the imaginary part of
the electron self-energy is given by
\begin{eqnarray}
Im[\Sigma^{c} (k_{F}^{c}, \omega + i\delta)] &=& - \frac{1}{\pi }
\int_{-\infty}^{\infty} \frac{d q_{\parallel}}{2\pi}
\int_{-\infty}^{\infty} \frac{d q_{\perp}}{2\pi}
\int_{-\infty}^{\infty} d \nu \int_{-\infty}^{\infty} d \nu'
\delta(\omega - \nu + \nu') [n(\nu) + f(\nu')] \cr &&\hspace{3cm}
\times \frac{\phi + \gamma_{b} \frac{\nu}{q_{\perp}}}{ \Bigl( \nu
+ \mu - \frac{q_{\perp}^{2}}{2m_{h}} -
\gamma_{b}\frac{\phi}{q_{\perp}} \Bigr)^{2} + \Bigl( \phi +
\gamma_{b} \frac{\nu}{q_{\perp}} \Bigr)^{2} } \frac{\phi}{(\nu -
v_{F}^{\psi} q_{\parallel})^{2} + \phi^{2}}
\cr
&& -\, \frac{1}{\pi } \int_{-\infty}^{\infty} \frac{d
q_{\parallel}}{2\pi} \int_{-\infty}^{\infty} \frac{d
q_{\perp}}{2\pi} \int_{-\infty}^{\infty} d \nu
\int_{-\infty}^{\infty} d \nu' \delta(\omega - \nu + \nu') [n(\nu)
+ f(\nu')] \cr
&&\hspace{3cm}\times \frac{-\phi + \gamma_{b} \frac{\nu}{q_{\perp}}}{
\Bigl( \nu + \mu - \frac{q_{\perp}^{2}}{2m_{h}} +
\gamma_{b}\frac{\phi}{q_{\perp}} \Bigr)^{2} + \Bigl( -\phi +
\gamma_{b} \frac{\nu}{q_{\perp}} \Bigr)^{2} } \frac{-\phi}{(\nu -
v_{F}^{\psi} q_{\parallel})^{2} + \phi^{2}}\, ,
\end{eqnarray}
where the presence of the Polyakov-loop parameter is the key feature.

It is straightforward to integrate over $q_{\parallel}$, which yields
\begin{eqnarray}
&& Im[ \Sigma^{c} (k_{F}^{c}, \omega + i\delta)] \cr
& =& -
\frac{1}{2\pi v_{F}^{\psi}} \int_{-\infty}^{\infty} \frac{d
q_{\perp}}{2\pi} \int_{-\infty}^{\infty} d \nu
\int_{-\infty}^{\infty} d \nu' \delta(\omega - \nu + \nu') [n(\nu)
+ f(\nu')] \frac{\phi + \gamma_{b} \frac{\nu}{q_{\perp}}}{ \Bigl(
\nu + \mu - \frac{q_{\perp}^{2}}{2m_{h}} -
\gamma_{b}\frac{\phi}{q_{\perp}} \Bigr)^{2} + \Bigl( \phi +
\gamma_{b} \frac{\nu}{q_{\perp}} \Bigr)^{2} }   \cr
&+& \frac{1}{2\pi v_{F}^{\psi}} \int_{-\infty}^{\infty} \frac{d
q_{\perp}}{2\pi} \int_{-\infty}^{\infty} d \nu
\int_{-\infty}^{\infty} d \nu' \delta(\omega - \nu + \nu') [n(\nu)
+ f(\nu')] \frac{-\phi + \gamma_{b} \frac{\nu}{q_{\perp}}}{ \Bigl(
\nu + \mu - \frac{q_{\perp}^{2}}{2m_{h}} +
\gamma_{b}\frac{\phi}{q_{\perp}} \Bigr)^{2} + \Bigl( -\phi +
\gamma_{b} \frac{\nu}{q_{\perp}} \Bigr)^{2} }\, .
\end{eqnarray}
It is essential to notice that the background potential is given
by $\phi(T) \approx \frac{\pi}{2} T$ in the confinement phase
because the Polyakov-loop parameter vanishes, i.e., $\cos \beta
\phi \approx 0$. Then, the potential $\phi$ has the same scaling
with the frequency. It is straightforward to see that the
imaginary part of the electron self-energy follows the typical
expression of $z = 3$ quantum criticality when $T \gg |\mu|$
because the holon excitations are effectively gapless in this
regime. As a result, we have $Im[\Sigma^{c} (T)] \propto T^{2/3}$.
On the other hand, the holon fluctuations become gapped in $T \ll
|\mu|$, and the self-energy corrections from these fluctuations
can be neglected, which gives rise to the typical Fermi liquid
form $Im[ \Sigma^{c} (T)] \propto T^{2}$.


\begin{thebibliography}{99}
\bibitem{Underdoped_Phase} The under-doped region at
intermediate temperatures is characterized by $U_{ij}^{SF} = -
\sqrt{\chi^{2} + \Delta^{2}} \tau_{3} \exp[i(-1)^{i_{x} + i_{y}}
\Phi \tau_{3}]$ with $\Phi = \tan^{-1}\Bigl( \frac{\Delta}{\chi}
\Bigr)$, $\langle h_{i} \rangle = 0$, and $a_{i0}^{k} = 0$, which
is known as the staggered flux (SF) phase, where the spinons have
the Dirac spectrum due to the staggered internal flux $\Phi$
whereas coherent electrons are not seen as the SM phase. Because
the electron spectrum exhibits its spectral gap except for Dirac
points, this SF state is identified with the so-called pseudogap
phase in high T$_{c}$ cuprates. Superconductivity arises from the
condensation of the holons $\langle b_{i1} \rangle \not= 0$ and
$\langle b_{i2} \rangle = 0$ due to $i a_{i0}^{k} \not= 0$ in the
SF phase.
\bibitem{SF_vs_SM} We would like to emphasize that the quadratic
dispersion for both spinon and holon dynamics is not valid in the
underdoped region, where the stable mean-field phase turns out to
be the SF phase, which gives rise to relativistic dispersions to
both fermions and bosons. As a result, an effective field theory
should be given by a relativistic SU(2) gauge theory except for
the holon sector, where its small chemical potential breaks the
relativistic invariance slightly. The chemical potential does not
arise in the spinon sector for the SF phase.
\bibitem{Data_fitting} If one sees our fitting curve carefully, the
theoretical line of the resistivity is not completely linear in
temperature, but slightly concave. Although the resistivity data
near the optimally doped region has been fitted by the T-linear
behavior, our demonstration shows that not only the T-linear form
but also a slightly concave expression can fit the data
qualitatively well.
\end{thebibliography}
\end{document}